\documentclass[12pt]{article} 
\usepackage{bbold}
\usepackage{cite}
\usepackage{graphicx}
\def \bea{\begin{eqnarray}} 
\def \beq{\begin{equation}}

\def \eea{\end{eqnarray}} 
\def \eeq{\end{equation}}

\def\lsim{\mathrel{\rlap{\lower3pt\hbox{$\sim$}}\raise2pt\hbox{$<$}}}
\def\gsim{\mathrel{\rlap{\lower3pt\hbox{$\sim$}}\raise2pt\hbox{$>$}}}

\begin{document} 
\begin{flushright}
EFI 15-18 \\
TECHNION-PH-2015-07 \\
June 2015 \\ 
\end{flushright} 
\centerline{\bf TRIPLE PRODUCT ASYMMETRIES IN $\Lambda_b$ AND $\Xi_b^0$ DECAYS}
\medskip
\centerline{Michael Gronau}
\centerline{\it Physics Department, Technion, Haifa 32000, Israel.}
\medskip 
\centerline{Jonathan L. Rosner} 
\centerline{\it Enrico Fermi Institute and Department of Physics,
  University of Chicago} 
\centerline{\it Chicago, IL 60637, U.S.A.} 
\bigskip

\begin{quote}                                 
The LHCb experiment is capable of studying four-body decays of the $b$-flavored
baryons $\Lambda_b$ and $\Xi_b^0$ to charmless final states consisting of
charged pions, kaons, and baryons.  We remark on the search in such modes for
CP-violating triple product asymmetries and for CP rate asymmetries relative
to decays involving charmed baryons.
\end{quote}

\leftline{\qquad PACS codes: 11.30.Er, 13.30.Eg, 14.20.Mr}

\section{Introduction}

In Ref.\ \cite{Gronau:2011cf} we presented a general discussion
for T-odd triple product (TP) asymmetries in four-body decays of strange,
charmed, and beauty mesons, focusing on genuine CP-violating asymmetries. 
Earlier studies of such asymmetries in $B$ meson decays to two charmless 
vector mesons have been made in 
Refs.~\cite{Valencia:1988it,Datta:2003mj,Datta:2011qz}.  The LHCb experiment
can extensively study four-body decays of the $b$-flavored baryons
$\Lambda_b$ and $\Xi_b^0$ into charged pions, kaons, and baryons
\cite{Neri}.  In the present paper we make some observations relevant to
the search for CP asymmetries in such decays.  Such asymmetries are expected
in charmless final states but not in charmed final states, which thus
provide an important control.

We discuss general features of $\Lambda_b$ and $\Xi_b^0$ decays to charmless
final states in Section \ref{sec:lb}.  Aspects of such decays relevant to CP
asymmetries are noted in Sec.\ \ref{sec:CP}, while we specialize to four-body
decays to charmless baryons and charged pions and kaons in Sec.\
\ref{sec:fourb}.  Resonant subsystems of four-body final states are discussed
in Sec.\ \ref{sec:sub}.  We conclude in Sec.\ \ref{sec:concl}.

\section{General features of charmless $\Lambda_b$ and $\Xi_b^0$ decays
\label{sec:lb}} 

Transitions of $b$-flavored baryons to charmless final states are governed
by two main subprocesses, which we will denote ``penguin" and ``tree".  The
penguin amplitudes effectively lead to a $b \to s$ transition when the decay
changes strangeness ($|\Delta S| = 1$) and $b \to d$ for $\Delta S = 0$.  The
$\Delta S = 0$ penguin amplitude $P$ is approximately $\lambda$ times that
($P'$) for ($|\Delta S| = 1$), where $\lambda = \tan \theta_C$ is the tangent
of the Cabibbo angle.  The tree subprocess $T$ is $b \to u \bar u d$ for
$\Delta S = 0$, while for $|\Delta S| = 1$ ($T'$) it is $b \to u \bar u s$,
with amplitude $\lambda$ relative to $T$.  From studies of $B$ meson decays
and low-multiplicity $\Lambda_b$ decays (see \cite{Gronau:2013mza} and
references therein) one can expect $|\Delta S| = 1$ processes to be
dominated by penguin amplitudes and $\Delta S = 0$ processes to be
dominated by trees.

\section{Aspects relevant to CP asymmetries \label{sec:CP}}

In order to observe a direct CP asymmetry, one needs two amplitudes with
different weak phases and different strong phases to interfere with one
another.  The asymmetry will be maximal when the amplitudes have comparable
magnitudes and relative weak and strong phases each as close as possible
to $90^\circ$.

For the tree and penguin amplitudes governing charmless $b$-flavored baryon
decays, the
relative weak phases are dominated by the large weak phase of the Cabibbo-%
Kobayashi-Maskawa matrix element $V_{ub}$: ${\rm arg}(V_{ub})\equiv -\gamma
\simeq =-67^\circ$ \cite{ckmf}.  The relative strong phases are not predictable
{\it a priori} but can be varied in multibody decay by varying the masses
of subsystems over the profiles of Breit-Wigner resonances.  The relative
magnitudes of tree and penguin ampitudes are in inverse proportion for
$|\Delta S| = 1$ and $\Delta S = 0$ amplitudes, giving the possibility of
a closer match between tree and penguin strength for one $\Delta S$ case or
the other.

The study of CP asymmetries in a proton-proton collider such as the CERN
Large Hadron Collider (LHC) is handicapped by the potentially unequal
production rates of particles and antiparticles.  For this reason (and
for cancellation of different detector sensitivities to particles and
antiparticles) it is useful to study CP-violating triple product asymmetries.
An asymmetry $A_T$ (as defined in Ref.\ \cite{Gronau:2011cf})
can arise without CP violation as a result of final-state interactions, but
should be equal to the asymmetry $\bar A_T$ for the corresponding antiparticle
decay if CP is conserved, so the difference
\beq
{\cal A}_T \equiv \frac{1}{2}(A_T - \bar A_T)
\eeq
provides a measure for CP violation.  Triple product asymmetries in two- and
three-body decays of polarized $\Lambda_b$ have been discussed in
Refs.~\cite{Kayser:1989vw,Bensalem:2000hq,Bensalem:2002pz, Bensalem:2002ys}. 

Another way of avoiding to a large extent uncertainties due to unequal
production rates of bottom baryons and antibaryons may be achieved by measuring
differences between CP rate asymmetries in charmless decays and in decays
involving charmed baryons. Differences in detector sensitivities to particles
and antiparticles may be minimized by choosing final states with identical
particles in charmless and charmed decays, taking into account their different
momenta.

\section{U-spin in four-body decays involving $\pi^\pm$ and $K^\pm$
\label{sec:fourb}}

We summarize accessible four-body charmless final states of $\Lambda_b$
involving protons, $\Sigma^+$ hyperons, charged pions, and charged kaons in
Table \ref{tab:dec}.  Also shown are final states of $\Xi_b^0 = bsu$.
We include $ \Sigma^+$ because it is related to the proton by a U-spin
reflection $d \leftrightarrow s$.  A similar transformation interchanges
$\Lambda_b$ and the lower-lying $\Xi_b^0$ (neglecting small configuration
mixing in the $\Xi_b^0$).  The $\Sigma^+$ decays to $n \pi^+$ (almost
impossible to identify) and $p \pi^0$ (requiring a detector to reconstruct
neutral pions).

\begin{table}
\caption{Four-body charmless final states involving a proton, a $\Sigma^+$, 
charged pions,
and charged kaons, in decays of $\Lambda_b = bud$ and $\Xi_b^0 = bsu$.
\label{tab:dec}}
\begin{center}
\begin{tabular}{c c c c} \hline \hline
  Decaying  & $|\Delta S|$ & Amplitudes &       Final         \\
  particle  &              &            &       state         \\ \hline
$\Lambda_b$ &      1       &  $T',P'$   & $p K^- \pi^+ \pi^-$ \\
            &              &            &   $p K^- K^+ K^-$   \\
            &               &           &   $\Sigma^+ \pi^- K^+ K^-$ \\
            &               &            &  $\Sigma^+ \pi^- \pi^+ \pi^-$ \\
            &      0       &   $T,P$    &  $p K^- K^+ \pi^-$  \\
            &              &            &  $p\pi^-\pi^+\pi^-$ \\
            &               &            &  $\Sigma^+ \pi^- K^+ \pi^-$ \\
$\Xi_b^0$   &      1       &  $T',P'$   &  $p K^- \pi^+ K^-$  \\
                   &               &                &  $\Sigma^+ \pi^-\pi^+ K^-$ \\
                   &               &                &   $\Sigma^+ K^- K^+ K^-$ \\
            &      0       &   $T,P$    & $p K^- \pi^+ \pi^-$ \\
            &              &            &   $p K^- K^+ K^-$   \\ 
            &              &            &  $\Sigma^+ \pi^- K^+ K^-$ \\
            &              &            &  $\Sigma^+ \pi^- \pi^+ \pi^-$ \\ 
            \hline \hline
\end{tabular}
\end{center}
\end{table}

CP (or T) violating triple-product asymmetries ${\cal A}_T$ may be formed from
each of these final states and four-body final states in corresponding
antibaryon decays.  
(CP violating triple-product correlations have already been 
investigated experimentally in charmed particle 
decays~\cite{delAmoSanchez:2010xj,Lees:2011dx,Aaij:2014qwa} where they
 are expected to be very small in the standard model~\cite{Kang:2009iy}.) 
 In the case of two identical particles (here, $K^- K^-$ or
$\pi^- \pi^-$) they are distinguished from one another by calling particle
number 1 the one with the higher momentum.

CP rate asymmetries for pairs of processes in which initial and final states
are obtained from each other by a U-spin reflection $d \leftrightarrow s$ have 
been shown to have equal magnitudes and opposite signs in the U-spin symmetry 
limit~\cite{Gronau:2000md,Gronau:2000zy,Bhattacharya:2013cvn}.  This property
has been confirmed experimentally in two-body $B^0$ and $B_s$ decays~%
\cite{Aaij:2013iua} and in phase-space-integrated three-body $B^+$ decays~\cite%
{Aaij:2013sfa}.  We will now show that, while similar relations hold also for
phase-space-integrated CP rate differences in four-body decays of bottom
baryons, such relations are not obeyed by the 
triple product CP asymmetries ${\cal A}_T$.

Consider, for instance, $\Lambda_b \to p(\vec p_1)K^-(\vec p_2)\pi^+(\vec p_3)\pi^-(\vec p_4)$ in the 
$\Lambda_b$ rest frame, $\Sigma_i \vec p_i = 0$. We define a T-odd triple product asymmetry
\beq
A_T \equiv \frac{\Gamma_{\Lambda_b}(C_T > 0) - \Gamma_{\Lambda_b}(C_T<0)}
{\Gamma_{\Lambda_b}(C_T > 0) + \Gamma_{\Lambda_b}(C_T<0)}
\equiv \frac{{\rm Cor}_{\Lambda_b}}
{\Gamma_{\Lambda_b}},
\eeq
where $C_T \equiv \vec p_1\cdot (\vec p_2 \times \vec p_3)$. In order to test CP violation we compare this asymmetry with a corresponding asymmetry in the CP conjugate process
$\bar \Lambda_b \to \bar p(-\vec p_1) K^+(-\vec p_2)\pi^-(-\vec p_3)\pi^+(-\vec p_4)$,
where the minus signs follows by applying parity to the three-momenta,
\beq
\bar A_T \equiv \frac{\Gamma_{\overline{\Lambda}_b}(C_T < 0) - 
\Gamma_{\overline{\Lambda}_b}(C_T> 0)}
{\Gamma_{\overline{\Lambda}_b}(C_T > 0) + 
\Gamma_{\overline{\Lambda}_b}(C_T<0)}
\equiv \frac{{\rm Cor}_{\overline{\Lambda}_b}}{\Gamma_{\overline{\Lambda}_b}}~.    
\eeq
Here $-C_T\equiv -\vec p_1\cdot (\vec p_2 \times \vec p_3)$ is the triple product of
 momenta for charge-conjugate particles. 

The difference 
\beq
{\cal A}_T \equiv \frac{1}{2}(A_T - \bar A_T)
\eeq
provides a measure for CP violation. A nonzero asymmetry ${\cal A}_T$,
\beq
\frac{{\rm Cor}_{\overline{\Lambda}_b}}{\Gamma_{\overline{\Lambda}_b}} \ne
\frac{{\rm Cor}_{\Lambda_b}}{\Gamma_{\Lambda_b}}~,
\eeq
may follow from a CP asymmetry in partial rates,
\beq
\Gamma_{\overline{\Lambda}_b} \ne \Gamma_{\Lambda_b}~,
\eeq
and/or from a CP asymmetry in triple-product correlations,
\beq
{\rm Cor}_{\overline{\Lambda}_b} \ne {\rm Cor}_{{\Lambda}_b}~.
\eeq

Now consider the decay $\Xi^0_b \to \Sigma^+(p_1)\pi^-(p_2)K^+(p_3)K^-(p_4)$ 
which is related to $\Lambda_b \to pK^-\pi^+\pi^-$ by a U-spin reflection, 
$d \leftrightarrow s$. 
Using the unitarity of the CKM matrix, 
\beq
{\rm Im}(V^*_{ub}V_{us}V_{cb}V^*_{cs}) = - {\rm Im}(V^*_{ub}V_{ud}V_{cb}V^*_{cd}),
\eeq
one may show that the two CP rate asymmetries have equal magnitudes 
and opposite signs in the U-spin symmetry limit~\cite{Gronau:2000zy,Bhattacharya:2013cvn}:
\beq
\Gamma_{\overline{\Lambda}_b} - \Gamma_{\Lambda_b} =
-[\Gamma_{\overline{\Xi}_b} - \Gamma_{\Xi_b}]~.
\eeq
A similar relation holds for corresponding triple product correlations,
\beq
{\rm Cor}_{\overline{\Lambda}_b} - {\rm Cor}_{{\Lambda}_b}
= -[{\rm Cor}_{\overline{\Xi}_b} - {\rm Cor}_{{\Xi}_b}]~.
\eeq
These two equations do not imply a relation between ${\cal A}_T(\Lambda_b)$ and 
${\cal A}_T(\Xi_b)$, namely between ${\rm Cor}_{\overline{\Lambda}_b}/\Gamma_%
{\overline{\Lambda}_b} - {\rm Cor}_{\Lambda_b}/\Gamma_{\Lambda_b}$ and 
${\rm Cor}_{\overline{\Xi}_b}/\Gamma_{\overline{\Xi}_b} -{\rm Cor}_{\Xi_b}/
\Gamma_{\Xi_b}$.
That is, separate opposite sign relations, (10) for CP asymmetries in triple
product correlations and (9) for corresponding decay rate asymmetries, do not
imply a similar relation for their ratios.  As mentioned, the requirement of
normalized CP violating triple product asymmetries follows from uncertainties
in relative production rates of baryons and antibaryons.  These uncertainties
may be largely avoided by studying CP rate asymmetries relative to
decays involving charmed baryons.

\section{Subsystems in four-body decays \label{sec:sub}}

Resonant subsystems in multibody final states offer the possibility of
controlling (or at least varying over a known range) the relative strong
phases of amplitudes, as long as the resonances are produced differently
by tree and penguin processes.  (See, for example, Refs.\ \cite{Atwood:%
1994zm,Eilam:1995nz}.)  We shall show this to be the case for the
processes of interest.

\begin{figure}
\begin{center}
\includegraphics[width=0.6\textwidth]{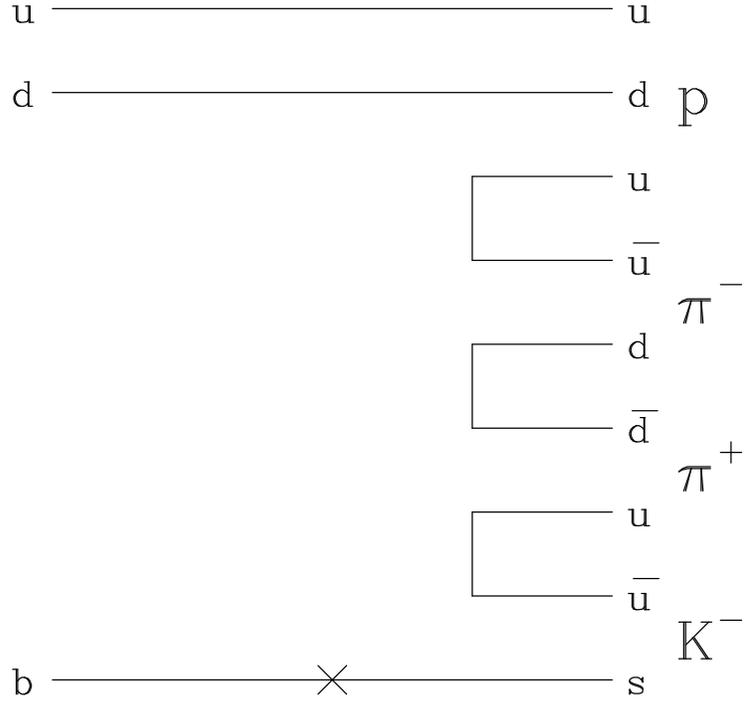}
\end{center}
\caption{Example of fragmentation graph for penguin $b \to s$ process in
$\Lambda_b \to p \pi^- \pi^+ K^-$.
\label{fig:frag}}
\end{figure}

Motivated by a picture of hadronization as due to quark-pair creation as
a QCD string stretches and undergoes fragmentation \cite{Andersson:1983ia},
one can draw graphs illustrating the formation of resonant subsystems
in four-body charmless decays of $\Lambda_b$ and $\Xi_b^0$.  Let us
take the $\Delta S = 1$ penguin process $b \to s$ in $\Lambda_b \to p K^-
\pi^+ \pi^-$ as an example.  One draws all possible ways of fragmenting
$sud$ into $p K^- \pi^+ \pi^-$, such that any two adjacent hadrons can
form a resonance.  Such a graph is shown in Fig.\ \ref{fig:frag}.
The results are shown in Table \ref{tab:frag}.

\begin{table}
\caption{Fragmentation of $\Lambda_b = bud \to sud$ into $p K^- \pi^+ \pi^-$
or permutations such that any two adjacent hadrons can form a resonance,
shown for the $b \to s$ penguin amplitude. The same eight orderings are allowed
for the $b \to s \bar u u$ amplitude.
\label{tab:frag}}
\begin{center}
\begin{tabular}{c c c c} \hline \hline
Final & \multicolumn{3}{c}{Resonance (example)} \\
state & 12 & 23 & 34 \\ \hline
$p \pi^- \pi^+ K^-$ & $N^{*0}$ & $\rho^0$ & $\bar K^{*0}$ \\
$\pi^- p \pi^+ K^-$ & $N^{*0}$ & $\Delta^{++}$ & $\bar K^{*0}$ \\
$K^- p \pi^+ \pi^-$ & $\Lambda^{*0}$ & $\Delta^{++}$ & $\rho^0$ \\
$K^- p \pi^- \pi^+$ & $\Lambda^{*0}$ & $N^{*0}$ & $\rho^0$ \\
$\pi^+ \pi^- p K^-$ & $\rho^0$ & $N^{*0}$ & $\Lambda^{*0}$ \\
$\pi^- \pi^+ p K^-$ & $\rho^0$ & $N^{*0}$ & $\Lambda^{*0}$ \\
$K^- \pi^+ p \pi^-$ & $\bar K^{*0}$ & $\Delta^{++}$ & $N^{*0}$ \\
$K^- \pi^+ \pi^- p$ & $\bar K^{*0}$ & $\rho^0$ & $N^{*0}$ \\ \hline \hline
\end{tabular}
\end{center}
\end{table}

The resonant subsystems one expects in this final state are thus
$N^{*0}$ (a generic $I=1/2$ or 3/2 nucleon resonance), $\rho^0$ or
any $\pi^- \pi^+$ resonance, $\bar K^{*0}$ and its excitations, and
any of numerous $K^-p$ resonances $\Lambda^{*0}$ such as $\Lambda(1520)$
\cite{PDG}.

\begin{figure}
\begin{center}
\includegraphics[width=0.6\textwidth]{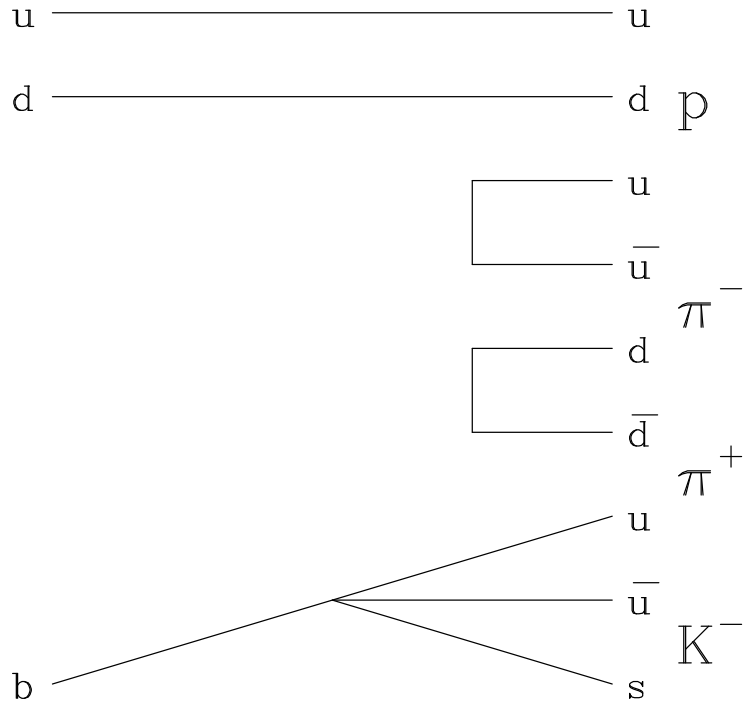}
\end{center}
\caption{Example of fragmentation graph for tree $b \to s \bar u u$ process in
$\Lambda_b \to p \pi^- \pi^+ K^-$.
\label{fig:tfrag}}
\end{figure}

We now consider resonant subsystem production by tree amplitudes.  In
this case the basic subprocess for $\Delta S = 1$ is $b \to u \bar u s$,
which requires one less light $q \bar q$ pair produced from the vacuum
than the penguin subprocess $b \to s$ to yield the same final state.
Consequently, the profile of resonance excitations by the tree amplitude
necessarily will differ from that of the penguin amplitude.  An example
for the final state $p \pi^- \pi^+ K^-$ is shown in Fig.\ \ref{fig:tfrag}.
%
In this case the most notable difference from Fig.\ \ref{fig:frag} is the
excitation of $\pi^+ K^-$ resonances.  This turns out to be so for all eight
orderings listed in Table \ref{tab:frag}.

In Ref.\ \cite{Gronau:2011cf} one could discuss 4-body spinless meson decays
with much greater specificity if they were dominated by quasi-two-body
channels such as a pair of vector mesons.  If quasi-two-body final states
dominate $\Lambda_b$ decays, helicity-amplitude decompositions may shed light
on CP-violating triple products.  Any of the (12) and (34) pairings in
Table \ref{tab:frag} could be expected to exhibit quasi-two-body behavior.
Natural sets of variables exist (e.g., \cite{Valencia:1988it,Dighe:1995pd})
for parametrizing such decays.

\section{Concluding remarks \label{sec:concl}}

We have discussed triple-product CP asymmetries in four-body decays of
$\Lambda_b$ and $\Xi_b^0$ to a proton or a $\Sigma^+$ hyperon, charged pions,
and charged kaons.  Decays involving a proton are most likely to be observed
and interpretable when the final-state hadrons experience resonant
substructure.  In that case, scanning across a Breit-Wigner resonance in the
effective mass of a two-body subsystem is guaranteed to produce a final-state
phase varying over an interval of nearly $180^\circ$.  In order for this phase
to contribute to a CP asymmetry, the interfering penguin and tree amplitudes
have to produce the resonance with different initial phases.  The different
production topologies for penguin and tree amplitudes strongly suggest this
will be the case. 

Decays involving a $\Sigma^+$ are related by a U-spin reflection to 
corresponding decays with a proton. Integrated CP rate asymmetries in these
pairs of processes are predicted to have approximately equal magnitudes but
opposite signs. Performing such direct tests at the LHC requires knowledge of
ratios of production rates for bottom baryons and bottom antibaryons
and of ratios of detector sensitivities for low-lying baryons and antibaryons.
Uncertainties due to unequal production rates of bottom baryons and antibaryons 
may be avoided to a large extent by measuring differences between CP rate 
asymmetries in charmless decays and in decays involving charmed baryons.

One can extend the present discussion to final states involving two baryons and
an antibaryon, such as $p \bar p p K^-$ and $p \bar p p \pi^-$.  Fragmentation
diagrams resemble those of Figs.\ \ref{fig:frag} and \ref{fig:tfrag}, except
that instead of a $d \bar d$ pair in the middle of the chain, one has an
antidiquark--diquark pair $[\bar u \bar d][ud]$.  The mass distributions are
likely to be dominated by low-mass $p \bar p$ enhancements (e.g., $X(1835)$
\cite{PDG}), as observed in $B$ decays~\cite{Aubert:2005gw}, so quasi-two-body
groupings $(p \bar p)(p K^-)$ or $(p \bar p)(p \pi^-)$ are likely to be useful.

\section*{Acknowledgments} We thank Nicola Neri for extensive discussions.
J.L.R.\ is grateful to the Technion for hospitality during the inception of
this work, which was supported in part by the United States
Department of Energy through Grant No.\ DE FG02 90ER40560.
M. G. wishes to thank the Munich Institute for Astro and Particle Physics for
its hospitality and support.


\begin{thebibliography}{99}

\bibitem{Gronau:2011cf} M.~Gronau and J.~L.~Rosner,
  Phys.\ Rev.\ D {\bf 84} (2011) 096013 [arXiv:1107.1232 [hep-ph]].
  
\bibitem{Valencia:1988it} G.~Valencia,
  Phys.\ Rev.\ D {\bf 39} (1989) 3339.
  
\bibitem{Datta:2003mj} A.~Datta and D.~London,
  Int.\ J.\ Mod.\ Phys.\  A {\bf 19} (2004) 2505 [arXiv:hep-ph/0303159].

\bibitem{Datta:2011qz} A.~Datta, M.~Duraisamy and D.~London,
  Phys.\ Lett.\  B {\bf 701} (2011) 357 [arXiv:1103.2442 [hep-ph]].

\bibitem{Neri} N. Neri, private communication.

\bibitem{Gronau:2013mza} M.~Gronau and J.~L.~Rosner,
  Phys.\ Rev.\ D {\bf 89} (2014) 037501 [arXiv:1312.5730 [hep-ph]].

\bibitem{ckmf} CKMfitter Collaboration, {\tt ckmfitter.in2p3.fr}.
 
\bibitem{Kayser:1989vw} B.~Kayser,
  Nucl.\ Phys.\ Proc.\ Suppl.\ {\bf 13} (1990) 487.

\bibitem{Bensalem:2000hq} W.~Bensalem and D.~London,
  Phys.\ Rev.\ D {\bf 64} (2001) 116003 [hep-ph/0005018].

\bibitem{Bensalem:2002pz} W.~Bensalem, A.~Datta and D.~London,
  Phys.\ Lett.\ B {\bf 538} (2002) 309 [hep-ph/0205009].

\bibitem{Bensalem:2002ys} W.~Bensalem, A.~Datta and D.~London,
  Phys.\ Rev.\ D {\bf 66} (2002) 094004 [hep-ph/0208054].
  
\bibitem{delAmoSanchez:2010xj}  
  P.~del Amo Sanchez {\it et al.} (BaBar Collaboration),
  Phys.\ Rev.\ D {\bf 81} (2010) 111103 [arXiv:1003.3397 [hep-ex]].
 
\bibitem{Lees:2011dx} J.~P.~Lees {\it et al.} (BABAR Collaboration),
  Phys.\ Rev.\ D {\bf 84} (2011) 031103 [arXiv:1105.4410 [hep-ex]].
  
\bibitem{Aaij:2014qwa} R.~Aaij {\it et al.} (LHCb Collaboration),
  JHEP {\bf 1410} (2014) 005 [arXiv:1408.1299 [hep-ex]].
  
  \bibitem{Kang:2009iy} See e. g.,
  X.~W.~Kang and H.~B.~Li,
  Phys.\ Lett.\ B {\bf 684} (2010) 137
  [arXiv:0912.3068 [hep-ph]];
  X.~W.~Kang, H.~B.~Li, G.~R.~Lu and A.~Datta,
  Int.\ J.\ Mod.\ Phys.\ A {\bf 26} (2011) 2523
  [arXiv:1003.5494 [hep-ph]].
  
  \bibitem{Gronau:2000md} M.~Gronau and J.~L.~Rosner,
  Phys.\ Lett.\ B {\bf 482} (2000) 71 [hep-ph/0003119].

\bibitem{Gronau:2000zy} M.~Gronau,
  Phys.\ Lett.\ B {\bf 492} (2000) 297 [hep-ph/0008292].
  
\bibitem{Bhattacharya:2013cvn} B.~Bhattacharya, M.~Gronau and J.~L.~Rosner,
  Phys.\ Lett.\ B {\bf 726} (2013) 337 [arXiv:1306.2625 [hep-ph]]. 
  
\bibitem{Aaij:2013iua} R.~Aaij {\it et al.} (LHCb Collaboration),
  Phys.\ Rev.\ Lett.\ {\bf 110} (2013) 221601 [arXiv:1304.6173 [hep-ex]].
  
  \bibitem{Aaij:2013sfa} R.~Aaij {\it et al.} (LHCb Collaboration),
  Phys.\ Rev.\ Lett.\ {\bf 111} (2013) 101801 [arXiv:1306.1246 [hep-ex]];
  Phys.\ Rev.\ Lett.\ {\bf 112} (2014) 011801 [arXiv:1310.4740 [hep-ex]];
  Phys.\ Rev.\ D {\bf 90} (2014) 112004 [arXiv:1408.5373 [hep-ex]]. 
  
\bibitem{Atwood:1994zm} D.~Atwood, G.~Eilam, M.~Gronau and A.~Soni,
  Phys.\ Lett.\ B {\bf 341} (1995) 372 [hep-ph/9409229].

\bibitem{Eilam:1995nz} G.~Eilam, M.~Gronau and R.~R.~Mendel,
  Phys.\ Rev.\ Lett.\ {\bf 74} (1995) 4984 [hep-ph/9502293].

\bibitem{Andersson:1983ia} 
  B.~Andersson, G.~Gustafson, G.~Ingelman and T.~Sjostrand,
  Phys.\ Rept.\ {\bf 97} (1983) 31.

\bibitem{PDG} K. A. Olive {\it et al.} (Particle Data Group),
  Chin.\ Phys.\ C {\bf 38} (2014) 090001.

\bibitem{Dighe:1995pd} A.~S.~Dighe, I.~Dunietz, H.~J.~Lipkin and J.~L.~Rosner,
  Phys.\ Lett.\ B {\bf 369} (1996) 144 [arXiv:hep-ph/9511363].
  
 \bibitem{Aubert:2005gw} B.~Aubert {\it et al.} (BaBar Collaboration),
Phys.\ Rev.\ D {\bf 72} (2005) 051101 [hep-ex/0507012];
Phys.\ Rev.\ D {\bf 76} (2007) 092004 [arXiv:0707.1648 [hep-ex]];
 M.~Z.~Wang {\it et al.} (Belle Collaboration),
 Phys.\ Rev.\ Lett.\ {\bf 92} (2004) 131801 [hep-ex/0310018];
 J.~T.~Wei {\it et al.} (Belle Collaboration),
 Phys.\ Lett.\ B {\bf 659} (2008) 80 [arXiv:0706.4167 [hep-ex]];
R.~Aaij {\it et al.} (LHCb Collaboration),
Phys.\ Rev.\ D {\bf 88} (2013), 052015 [arXiv:1307.6165 [hep-ex]].

\end{thebibliography}
\end{document}